\documentclass[journal=apchd5,manuscript=article]{achemso}
\usepackage{chemformula} 
\usepackage[T1]{fontenc} 
\usepackage{natbib}

\author{Louis Nicolas}
\affiliation{Laboratoire de Physique de l'Ecole normale sup\'erieure, ENS, Universit\'e PSL, CNRS, Sorbonne Universit\'e, Universit\'e Paris-Diderot, Sorbonne Paris Cit\'e, Paris, France.}

\author{Tom Delord}
\affiliation{Laboratoire de Physique de l'Ecole normale sup\'erieure, ENS, Universit\'e PSL, CNRS, Sorbonne Universit\'e, Universit\'e Paris-Diderot, Sorbonne Paris Cit\'e, Paris, France.}

\author{Paul Huillery}
\affiliation{Laboratoire de Physique de l'Ecole normale sup\'erieure, ENS, Universit\'e PSL, CNRS, Sorbonne Universit\'e, Universit\'e Paris-Diderot, Sorbonne Paris Cit\'e, Paris, France.}

\author{Cl\'ement Pellet-Mary}
\affiliation{Laboratoire de Physique de l'Ecole normale sup\'erieure, ENS, Universit\'e PSL, CNRS, Sorbonne Universit\'e, Universit\'e Paris-Diderot, Sorbonne Paris Cit\'e, Paris, France.}

\author{Gabriel~H\'etet}
\email{gabriel.hetet@lpa.ens.fr}
\affiliation{Laboratoire de Physique de l'Ecole normale sup\'erieure, ENS, Universit\'e PSL, CNRS, Sorbonne Universit\'e, Universit\'e Paris-Diderot, Sorbonne Paris Cit\'e, Paris, France.}

\title{Sub-GHz linewidths ensembles of SiV centers in a diamond nano-pyramid revealed by charge state conversion} 
   
\begin{document}

\begin{abstract}
Producing nano-structures with embedded bright ensembles of lifetime-limited emitters is a challenge with potential high impact in a broad range of physical sciences. In this work, we demonstrate controlled charge transfer to and from dark states exhibiting very long lifetimes in high density ensembles of SiV centers hosted in a CVD-grown diamond nano-pyramid.
Further, using a combination of resonant photoluminescence excitation and a frequency-selective persistent hole burning technique that exploits such charge state transfer, we could demonstrate close to lifetime-limited linewidths from the SiV centers.
Such a nanostructure with thousands of bright narrow linewidth emitters in a volume much below $\lambda^3$ will be useful for coherent light-matter coupling, for biological sensing, and nanoscale thermometry.
\end{abstract}


\section*{Introduction}

Photoluminescent defects in diamond are atom-like systems in the solid state that have unique photophysical properties. 
Most of them indeed feature stable and bright photoluminescence (PL) at room temperature, making them ideal single photon sources. In particular, the electronic spin of the negatively charged nitrogen vacancy (NV$^-$) defect can be manipulated at room temperature, used to probe magnetic and electric fields, as well as nearby nuclear spins, and as a temperature and strain probe \cite{Bernardi2017}. Another defect, the negatively charged silicon vacancy center (SiV$^-$) has also received considerable interest over the past decade.  One advantage of the SiV$^-$ center over the NV$^-$ center is that most of its photoluminescence concentrates in the zero-phonon line. 
Moreover, SiV$^-$ centers can also provide optical qubit states with long coherence times exceeding 10 ms at temperatures below 500 mK \cite{Sukachev2017}.  Another advantage of the SiV$^-$ center spectral lines is that they are almost insensitive to nearby charge fluctuations \cite{Rogers2014b}, due to their inversion symmetry,
which leads to a small inhomogeneous broadening. 

Using color centers within nanostructures and nanodiamond (NDs) is essential for many applications in sensing, labeling and quantum information. Using NDs also means more efficient collection of the photoluminescence because the light does not undergo total internal reflection in the diamond.
Producing ensembles of SiV$^-$ centers in diamond nanostructures while remaining close to the lifetime limit is however challenging because preserving the axial symmetry is a difficult task. 
Indeed, in NDs, strain breaks the axial symmetry, which makes the centers very sensitive to surface charge fluctuations and in turn increases the spectral diffusion.
There is however great progress in this direction. 
Single SiV$^-$ centers in nanodiamonds grown on iridium by chemical vapour deposition (CVD) have exhibited 20 GHz linewidths, limited by spectral diffusion \cite{Neu2013}. Better spectral properties have been reached for nanodiamonds synthetized by CVD on silicon (325 MHz)  \cite{Li2016}.
Single SiV$^-$ centers with lifetime limited linewidth (approx. 100 MHz) has also been achieved in diamond nanostructures using ion implantation \cite{englund_emp, Evans2016, Marseglia2018}.  
Last nanodiamonds grown via a high-pressure high-temperature method (HPHT) and subsequently treated in hydrogen plasma have exhibited very close to lifetime limited single SiV$^-$ centers (354 MHz) \cite{jantzen2016nanodiamonds}. 

The static spectral shifts due to strain are however typically very large in nanostructured diamonds. It can be turned into an advantage since single SiV$^-$ can be spectrally isolated, but this precludes reaching the lifetime limit with  ensembles of SiV$^-$ centers, which can significantly increase the photoluminescence rate. In this work, we use CVD grown low-strain nano-pyramids together with charge state conversion as a means to counteract this issue. 

For most defects in diamond, charge state conversion plays an important role. It can lead to an overall decay of the PL and can necessitate multicolor laser pulse sequences for repopulating the desired charged states. 
Conversely, charge state conversion can have advantages. It was for instance used to demonstrate detection of the NV$^-$ electronic spin state \cite{Shields2015} and lead to the development of ground-state depletion microscopy (GSD) \cite{NVGSD} and 
persistent hole burning \cite{Macfarlane, Redman92}. 
As in the case of NV centers, switching the charge state of SiV centers was also observed \cite{MerilesSiV} and used to manipulate single SiV$^-$ centers \cite{Evanseaau4691}. 

Here we first demonstrate optical and fatigue-free charge state control over a high density of SiV centers located at the apex of a diamond nano-pyramid. 
We then use resonant photoluminescence excitation (PLE) in conjunction with persistent hole burning {\it via} charge state conversion, to show close to lifetime-limited emission of thousands of bright emitters in a volume below $\lambda^3$. 

\section*{Samples and experimental set up}

\begin{figure}[ht!!]
\centerline{\scalebox{0.15}{\includegraphics{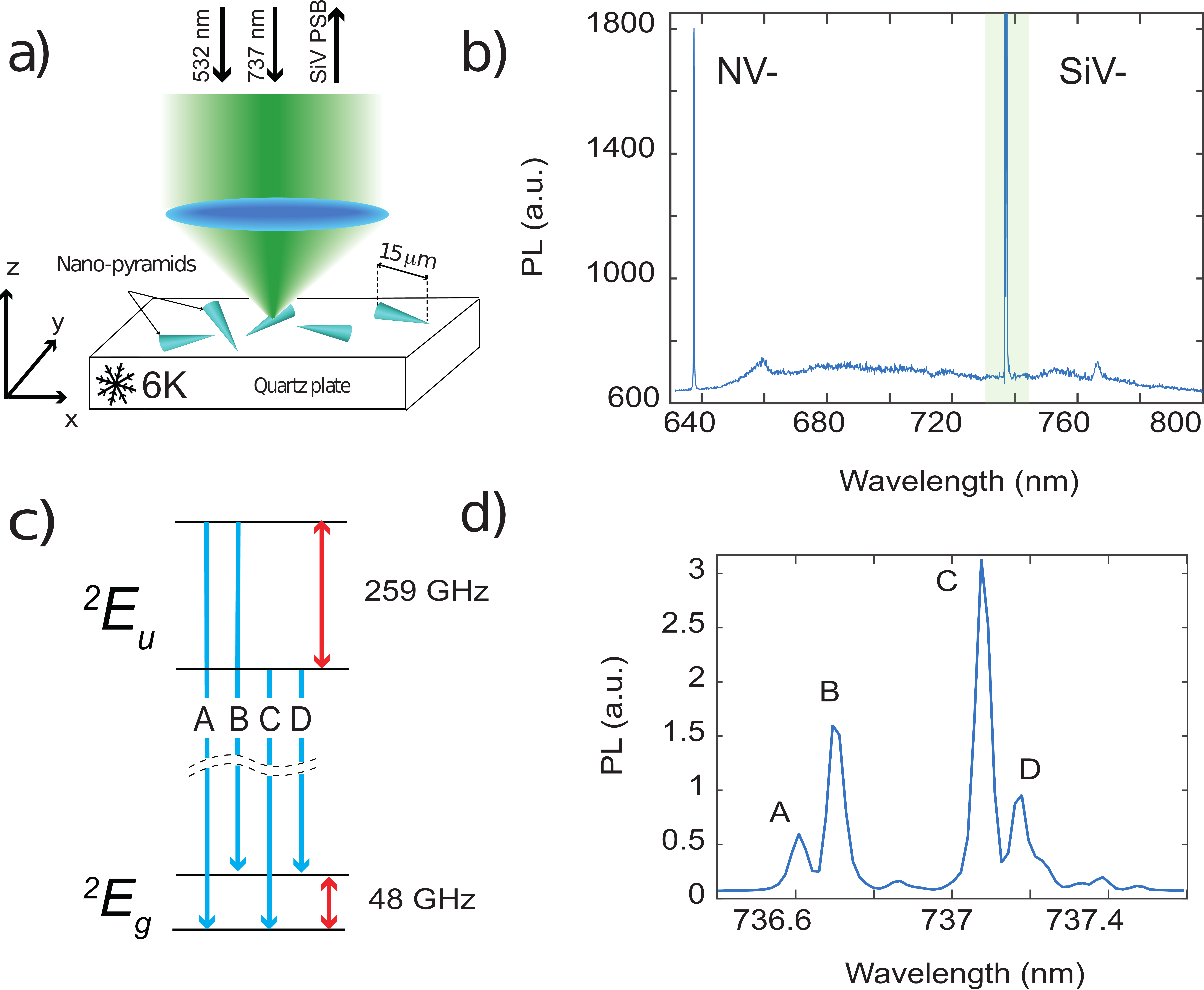}}}
\caption{a) Schematics of the confocal microscope set-up, showing nano-pyramids deposited on a quartz plate, and the two employed lasers at 532 nm and 737 nm. 
b) Optical spectrum of the photoluminescence (PL) emitted by the pyramids in the 630 to 800 nm range. 
c) Level scheme of the SiV$^-$ center. d) PL spectrum close to the SiV center, showing the four characteristic ZPLs under green excitation. The smaller other peaks correspond to the photoluminescence from other silicon isotopes.}\label{setup}
\end{figure}

The samples under study are CVD grown Artech Carbon AFM tips.
These samples were investigated in Ref.~\citenum{Nelz2016, Nicolas2018, choi2018}. Ensembles of SiV$^-$ centers are located at the ten nanometer radius apex of the tip and exhibit an inhomogeneous broadening that is less than 10 GHz at 4K \cite{Nicolas2018}.
We have two different types of samples at our disposal : pyramids containing nitrogen impurities (with a concentration of around 10 ppm  \cite{Nelz2016}) and pyramids where no nitrogen was detected. Both contain SiV centers mostly at the apex \cite{Nicolas2018}. In the present study, we mainly focus on the first sample type because of its peculiar properties.
The pyramids are deposited on quartz or silicon substrates as depicted in Fig. \ref{setup}-a). To study their optical properties, we use the same homebuilt confocal microscope at cryogenic temperatures as in Ref.~\citenum{Nicolas2018} with the addition of a laser tuned to the 737 nm zero-phonon line.

In the samples with nitrogen, a significant fraction of the photoluminescence is emitted by NV centers.
An optical spectrum measured at the apex of one pyramid is shown in Fig. \ref{setup}-b). It has been obtained at 6 K, using 4 mW of green laser excitation focused on the sample by a 0.25 numerical aperture objective, and taken on a 1200 lines/mm grating spectrometer. 
The zero-phonon line (ZPL) of both the NV$^-$ and SiV$^-$ centers are observed at 637 and 737 nm respectively. 
The spectrum also features the phonon sidebands (PSB) of both centers. The NV$^-$ center PSB emission ranges from 645 nm to 730 nm. 
while it ranges from 740 to 765 nm for in the SiV$^-$ center and also features the local silicon oscillation mode at 763 nm \cite{Neu2011}.
Due to a similar symmetry between the SiV$^-$ ground and excited states, the electron coupling strength to the phonons is much weaker than for the NV$^-$ center, so the relative area and spectral spread under their PSB is much smaller. 

Using a band-pass filter (730-770 nm) around the SiV$^-$ ZPL, we found photoluminescence count rates of around 1 million counts per seconds with 1 mW of green laser light.
This very high count rate implies a high SiV concentration, considering the fairly small numerical aperture (NA=0.25) that we use in the experiment and the fact that most of the emission is funneled through the diamond tip\cite{choi2018}. 

A zoom on the SiV$^-$ ZPL is plotted in Fig.\ref{setup}-d). The four characteristic transitions of the SiV$^-$ centers are observed together with smaller peaks corresponding to the photoluminescence from other silicon isotopes.
Indeed, their positions are consistent with the ZPL of the isotopes of silicon and their relative heights and natural abundance \cite{Dietrich2014}.
The two ground and excited states are splitted by 48 GHz and 259 GHz respectively due to spin-orbit coupling. 
Observing such a bulk-like SiV$^-$ spectral response with ensembles of SiV$^-$ centers already underlines the very small strain in the nano-pyramid. 

Photoluminescence excitation using a laser that is scanned about the ZPL is an efficient method to obtain an even higher spectral resolution. In PLE experiments with SiV$^-$ centers, a resonant laser beam at 737 nm is scanned close to the SiV$^-$ ZPL, and the PL is collected on the phonon side band (PSB) after filtering the laser light using a 750 nm longpass filter. 


\begin{figure}[ht!!]
\centerline{\scalebox{0.15}{\includegraphics{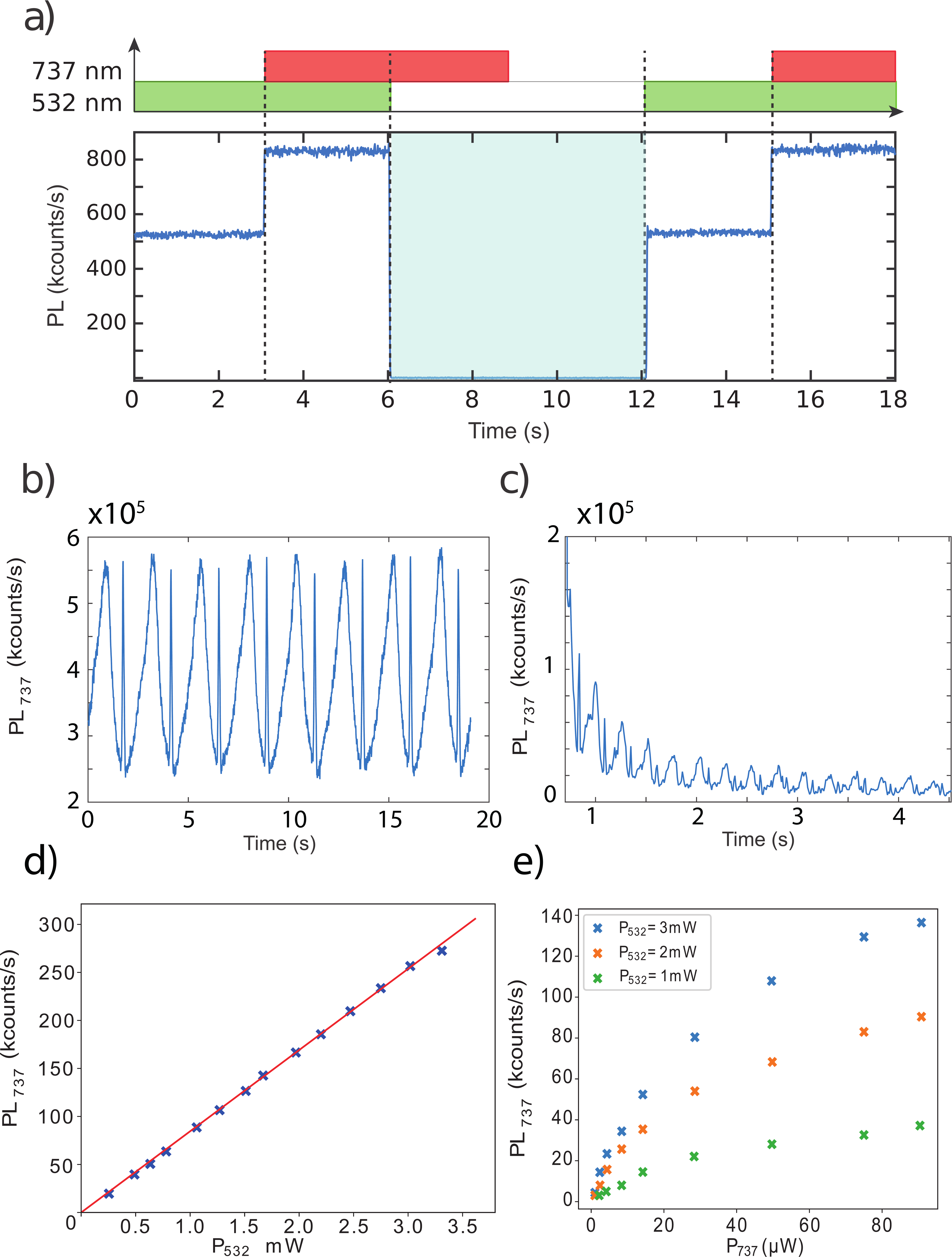}}}
\caption{a) Laser sequence used to reveal the trapping of the SiV$^-$ center into a dark state. The power of the green and resonant lasers are $P_{532}$= 4.3 $\mu$W and $P_{737}$= 9.5 $\mu$W. 
b) PL count rate on the SiV phonon-sideband, as the resonant laser is repeatedly scanned about the transition C of a sample containing no detectable nitrogen. 
c) Similar scan performed on a sample containing nitrogen. The green laser was turned off at the beginning of the scan sequence. 
d) PL count rate as a function of green laser power (with the resonant laser off) and linear fit to the data (plain line). e) PL count rate due to the resonant excitation as a function of the resonant laser power, for three different green laser powers.
}\label{modele}
\end{figure}

\section*{Quenching of the Photoluminescence}

In our first attempts to observe PLE on the pyramids which contain nitrogen, we could not detect PL on the PSB under resonant excitation only. However, when the green laser was present, the resonant laser increased the PL count rate. 
Fig. \ref{modele}-a) shows a typical time trace where the PL count rate in the 750 and 770 nm range is plotted 
for 18 s during which green and resonant lasers are turned on and off and kept at optical powers of  4.3 mW and 9.5 $\mu$W respectively. Their wavelengths are 532 nm and 737 nm respectively.
When the green laser is turned on, a count rate of 520 kcounts/s is measured. This count rate is stable for hours. We then turn on the resonant laser at a time t = 2.6 s which increases the count rate to 830 kcounts/s, with also similar long-term stability. When turning off the green laser, at a time t = 6 s however, the photoluminescence of the SiV$^-$ center is completely suppressed.  
We attribute this quenching of the photoluminescence to the trapping into a dark state of the SiV$^-$ centers.
This dark state can be switched again back on by applying a laser at 532 nm. We repeated the same procedure and measured the same photoluminescence rates by applying the green and red laser sequence again. Crucially, this effect is only seen when using pyramids that contain nitrogen. 
This is seen in Fig. \ref{modele}-b), where the PL count rate in the 750 and 770 nm range is detected as the resonant laser is repeatedly scanned about the transition C of a sample containing no detectable nitrogen (at a laser power of 360 nW). This measurement was done without using additional green laser and with a saw-tooth scan with a 10 GHz frequency span. 
The sharp peaks in between each SiV line is PL during the rapid scan when the laser frequency returns to its initial value. It can be noticed that the maximum of the signal stays at the same level (about 550 kcounts/s).
Fig. \ref{modele}-c) shows a similar scan performed on a sample containing nitrogen at a laser power of 250 nW. This time, the green laser was kept on and then suddenly turned off at the beginning of the scan sequence. It can be seen here that the PL maxima decay on second time scales.
This observation points towards a decisive role played by nitrogen impurities.

To gain further information about the nature of the dark state, we will study the PL rate evolution under different conditions. 

\section*{Charge State Transfer}

We record the PL count rates as a function of the green and red laser powers. Fig.~\ref{modele}-d) shows the PL count rate as a function of the green laser power for up to 5 mW, in the absence of red laser.  The linear dependence of the PL with respect to the green laser demonstrates that the SiV$^-$ centers optical transitions are not saturated. Fig.~\ref{modele}-e)
shows the PL count rate as a function of the resonant laser power from 0 to 90 $\mu$W, with the green laser power set to 1 mW, 2 mW and 3 mW (see inset). The plotted rate  (PL$_{737}$) is the total PL count rate with the PL due the green laser only subtracted off. 
It thus corresponds to the PL solely due to the resonant laser excitation. 
We observe that the maximum value of PL$_{737}$ depends on the green laser power : the higher the green laser power, the faster the SiV centers are converted back to their bright states. 
The inflection of the curves is not due to the saturation of the optical transition but to the competition between the trapping into the dark state and the recovery of the bright state.

\begin{figure}[ht!!]
\centerline{\scalebox{0.12}{\includegraphics{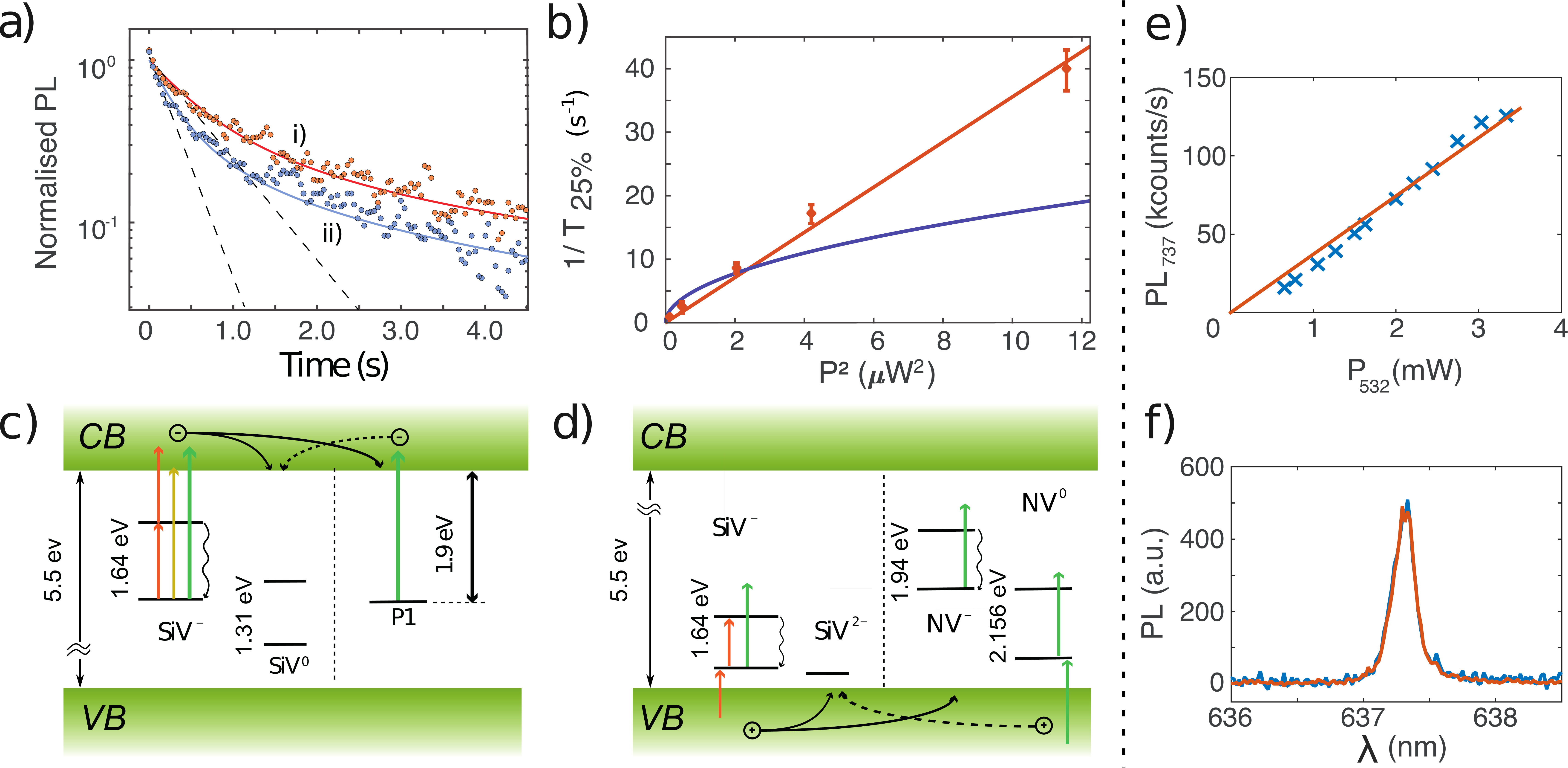}}}
\caption{
a) PL count rate as a function of time when the resonant laser is turned on at $t=$0 s for two different laser powers: i) P$_{737}$= 347 nW, ii) P$_{737}$= 683 nW.
b) Characteristic decay rate as a function of the square of the power of the resonant laser. In red, the linear fit as a function of the square of the laser power highlights the two-photon behavior. The blue line is a fit using a linear dependence with laser power.
c) Energy level scheme explaining the charge state conversion under the hypothesis that involves conduction band electrons. d) Energy level scheme explaining the charge state conversion under the hypothesis that involves holes. e) PL count rate due to the resonant laser as a function of the green laser power. The resonant laser power is set to 90 $\mu$W. f) Spectra of the NV$^-$ ZPL under green excitation (1 mW) with (in blue) and without (in red) 90 $\mu$W resonant excitation.}
\label{time}
\end{figure}

To better understand the phenomenon, we monitor the PL count rate after turning on the resonant laser and without green laser light. Fig.~ \ref{time}-a) shows the PL rate as a function of time for two resonant laser powers : P$_{737}$= 347 nW and P$_{737}$=683 nW . After a green illumination that prepares the SiV centers in the negatively charged state, the resonant laser is turned on at t = 0 s. The PL decays faster as the resonant laser power is larger on timescales of about 1 s. 
This means that the dynamics is not governed by the SiV$^-$ orbital states, the latter typically taking place on nanosecond timescales.
Importantly, no plateaus in the decay curve are observed which highlights the very high SiV center concentration. 
The solid lines are fits using the model that we describe in the "Persistent Hole Burning" section. 

Let us note that the initial PL count rate (at t=0 s) does not depend on the time during which the emitters have been let in the dark after the extinction of the green laser. We checked that this is the case even when waiting for hours. We also observe no PL when we turned on the resonant laser after an hour long period without any illumination once the emitters are trapped in the dark state. We can thus conclude that bright and dark states are stable over hours.
Considering the timescales of the PL decay and the long-term stability of the dark and bright states, the dark state is then likely to be another charge state of the SiV center.

Charge conversion processes often occur through photoionisation. This is indeed a common mechanism in diamond, which has been extensively studied with the NV$^-$ center \cite{NVGSD, Aslam2013, Siyushev2013, MerilesNV, Tallaire, Chen2017}.  As we will see, SiV centers reveal very similar behavior.
Previous studies on NV centers show that the NV$^-$ centers can be ionized under resonant excitation (637 nm) via a two-photon process. 
First, a photon promotes the NV$^-$ in its excited state and a second photon brings the electron to the diamond conduction band. This process forms NV$^0$ states plus one electron which may be trapped by another impurity nearby. 
The NV$^0$ center, the neutral NV form, can then be converted back into the NV$^-$ under green excitation : First, a photon brings the NV$^0$ in its excited state. Then, another photon promotes a valence electron (coming from the deep lying NV$^0$ $a_1$ orbital) to its fondamental state. The hole then migrates away from the center which is now in its negatively charged state.
Experimental studies have shown that only one photon can be enough for these two charge transfer processes when the centers are close to the surface \cite{MerilesNV}. \textit{Ab initio} calculations explain it by the presence of image and acceptor states very close to the surface \cite{kaviani2014proper}. During the process, electrons and holes are produced and are captured by the impurities, which stabilizes the charge state. Substitutional nitrogen centers, denoted as P1 centers, can play this role. Two stable P1 species exist: neutral and positively charged (P1$^0$ and P1$^+$). They can both accept electrons from the conduction band \cite{ulbricht2011single} and the neutral P1 can also efficiently capture holes from the valence band \cite{PAN1990}.

The charge transfer dynamics of the SiV$^-$ center is however still being investigated. 

In the literature, two different hypotheses involving two different SiV dark states can be found \cite{MerilesSiV,Gali2013,Thiering} and are depicted in Fig.~\ref{time}-c) and d).

\subsubsection*{Hypothesis 1: exchange of electrons}
The first hypothesis relies on the exchange of electrons via the conduction band : if the energy levels of the SiV$^-$ center lie close enough to the conduction band, an electron already in the SiV$^-$ excited state can be promoted to the conduction band under resonant excitation. Thus, the SiV$^-$ is converted into SiV$^0$. Then, the electron diffuses and is captured by impurities. If only SiV are present in the diamond, the electron is captured by the SiV$^0$ which returns to its charged state. Indeed, it is admitted that SiV$^0$ can recombine with an electron of the conduction band. 
That could explain why the PL is stable under resonant excitation when no nitrogen is present, as we experimentally checked. If nitrogen is present in the diamond however, the electron can be trapped by a positively charged substitutional nitrogen defect (P1 center) to form a neutral P1 center. 
This defect can further be photoionised using a more energetic excitation, such as with green light, and the electron promoted to the conduction band can be captured by a SiV$^0$ center which is converted back to the bright state\cite{PAN1990, Jones2009}. Fig.~\ref{time}-c) shows a depiction of this process. This hypothesis agrees with recent experimental studies \cite{MerilesSiV}.

\subsubsection*{Hypothesis 2 : exchange of holes}
The second hypothesis is based on the exchange of holes {\it via} the valence band.
This time the SiV$^-$ energy levels are supposed to lie close to the valence band, as some \textit{ab initio} calculations predict \cite{Gali2013,Thiering}. When the SiV$^-$ is in its excited state, an electron can be promoted from the valence band to its ground state under resonant excitation. This process converts the SiV center into its SiV$^{2-}$ state (the dark state under this hypothesis) and leaves a hole in the valence band. The hole can diffuse and can be captured back by the SiV$^{2-}$ but also by nitrogen related defects when present (NV$^-$ and P1$^0$ centers). Holes can be created again thanks to green excitation which converts NV$^0$ centers to NV$^-$ allowing the recovery of the SiV$^-$ photoluminescence. One can note that green light does not provide enough energy to create holes in the valence band when directly converting P1$^{+}$ center into P1$^0$ \cite{Jones2009} so only NV centers can play a role in this hypothesis. 

These two hypotheses are consistent with the need for nitrogen impurities in the dark state trapping process. They also both involved a two-photon process for charge state conversion.

Let us now focus on the rate of the state conversion process further, so as to verify the two-photon nature of the conversion.
Fig. \ref{time}-a) shows the photoluminescence count rate as a function of time on a log scale, for two different resonant laser powers P$_{737}$ : 347 nW and 683 nW for traces i) and ii) respectively. The decay process features two long timescales, a shorter one, in the hundreds of millisecond range, and a longer one, in the tens of second range.
These two time-scales are the result of excitation of an inhomogeneous class of emitters :
the emitters which are exactly at resonance are trapped into the dark state faster than ones which are detuned.
The short (resp. long) times scale thus correspond to the population dynamics of resonant (resp. off resonant) emitters,
which is explained by the model described below.

The characteristic faster decay rate is also plotted as a function of the square of the power in Fig.~\ref{time}-b). The decay rate plotted on the $y$ axis is the inverse of the duration needed to lower the PL by 75\%.
Since the optical transition of the SiV$^-$ centers is not saturated at these power levels (as shown in Fig.~\ref{modele}-c), a linear power dependency of the decay rate would be associated with single photon process, whereas quadratic dependency would reveal a two-photon process. The red line is a linear fit of this decay curve as a function of the square of the laser power. The blue line is a tentative fit that uses a linear dependence with laser power.  A fit using a two-photon process fits better our data. 
Note that this two-photon process is not seen on every pyramids however. Around 75\% of the dozen of studied pyramids show this behavior, the other show a linear evolution with laser power. 
It can be explained by the presence of surface defects as already observed with NV centers \cite{MerilesNV,kaviani2014proper}.

To discriminate the two above hypotheses, we have done several measurements. We first intended to observe the PL of the SiV$^0$ centers  (which exhibits a ZPL at 946 nm) after having converted the SiV centers into their dark state. We did not succeed in doing so using laser excitations at wavelengths of 532, 737, 800 and 935 nm.  This can be both the result of our poor detection efficiency at 946 nm and due to the presence of a state that decays non-radiatively.

We have also tried to measure a change in the NV$^-$ population to test the second hypothesis. Indeed, if resonant excitation of the SiV$^-$ centers produces holes, the NV$^-$ centers should be converted into NV$^0$ after hole capture. Fig.~\ref{time}-e) shows the PL count rate as a function of green laser power in the presence of the resonant laser set to relatively large powers ($P_{737}=90\mu$W). This count rate is directly linked to the number of SiV centers in the bright state. It shows that, at low green laser power, a significant portion of SiV centers are in the dark state which should result in a modification of the number of NV centers in the negatively charged state in the presence of the resonant laser.
Fig.~\ref{time}-f) shows the NV$^-$ center ZPL spectra under 1 mW of green laser excitation with (blue spectrum) and without (red spectrum) resonant excitation. No change in the NV$^-$ center population is observed, which tends to favor the first hypothesized charged state conversion mechanism \cite{Neu2012a,MerilesSiV}.

The PL rate plotted in Fig.~\ref{time}-e) is proportional to the population of the bright state as
the optical transition is not saturated by both lasers. The linear fit
thus implies that the transfer from the dark state to the bright SiV state
requires only one green photon which is consistent with a photoionisation of the $P_1$ center. However,
because we use a nanostructure, we cannot exclude that the
recombination process occurs through the creation of holes and from the
conversion of NV$^{0}$ centers into NV$^{-}$ centers
\cite{MerilesNV,kaviani2014proper}. 

It is important to note that, under both hypotheses, two conditions are required to observe the total loss of SiV$^-$ PL: the concentration of nitrogen impurities has to be high enough to capture all the charges coming from the photoionization of the SiV$^-$ centers and the Fermi level has to be appropriately pinned in order to enable such charge exchange \cite{Rose60,collins2002fermi}.
To illustrate it, let us assume that the first condition is fullfilled and suppose that the energy levels of the SiV$^-$ centers are close to the conduction band. If the Fermi level is pinned high enough, SiV centers naturally exist in their negatively charged states and P1 centers are in their neutral or even in their negatively charged states. They cannot capture the electrons resulting from the photoionisation of the SiV$^-$ centers. Consequently, they can be stable under resonant excitation. Similarly, in the second hypothesis which implies energy levels of the SiV$^-$ center close to the valence band, a too low Fermi level would lead to the stabilization of the SiV$^-$ because no impurities could accept holes coming from the SiV$^-$ $\rightarrow$ SiV$^{2-}$ conversion. Artificially shifting the Fermi level of our samples thanks to electrodes could thus discriminate one hypothesis from the other \cite{MerilesSiV,grotz2012charge}.

Having observed and analyzed this charge transfer mechanism and the role of the green laser in the charge reload mechanism, we now come back to the resonant PLE. 

\section*{Resonant Photoluminescence Excitation}

\begin{figure}[ht!!]
\centerline{\scalebox{0.1}{\includegraphics{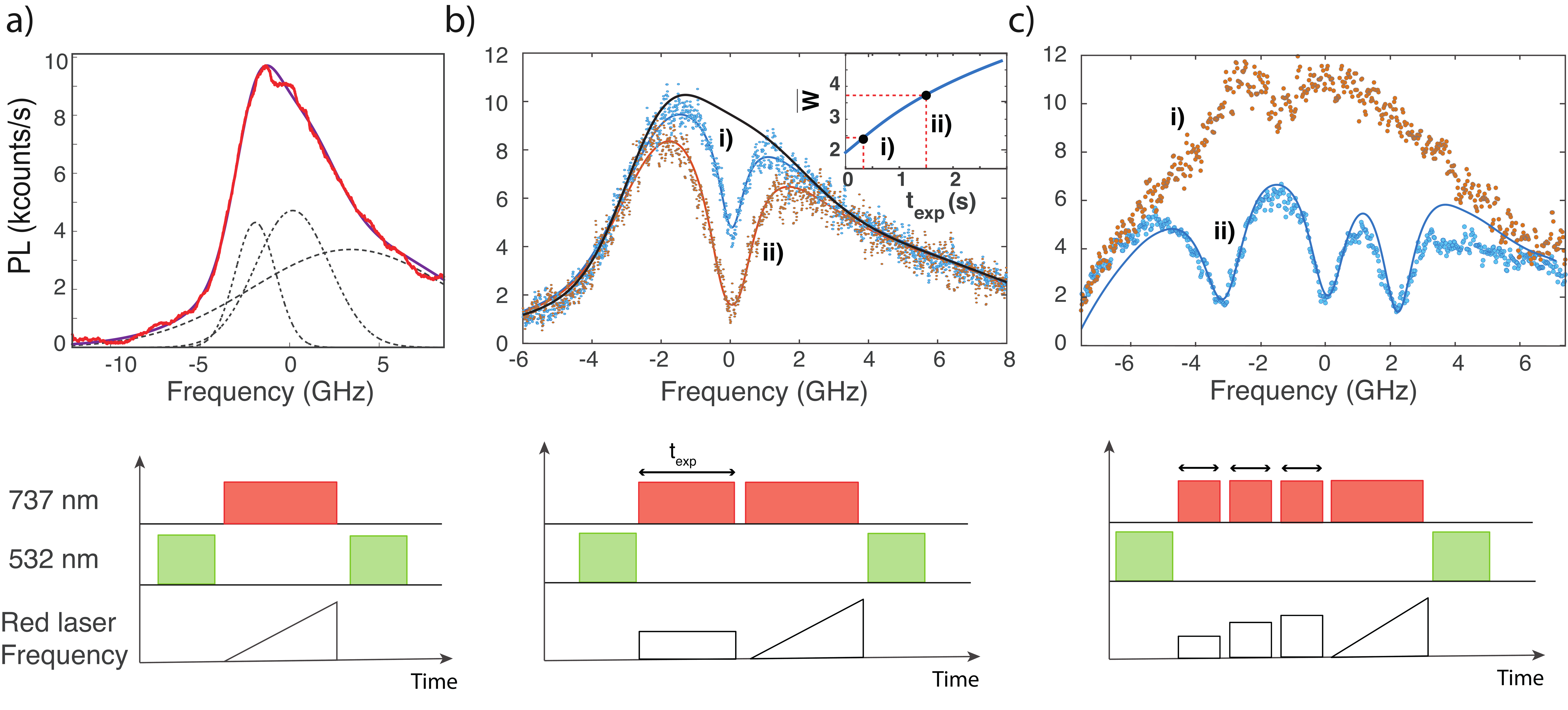}}}
\caption{a) Averaged PLE measured by scanning the resonant laser at 737 nm around transition C, in the presence of green laser light between scans.
b) PLE spectra after three different exposure times at 6K : in black without hole burning, in blue $t_{exp}=300$ ms and in orange, $t_{exp}=1500$ ms. The dots represent raw data and the continuous lines are fitted data. Here $P_{res}=0.877\mu W$. The inset shows the width $\overline{W}$ of the holes inferred from the fit and normalized to the homogeneous linewidth $\Gamma_e$ ($\overline{W}=\Gamma_h/\Gamma_e$). Notice the factor of two between the hole width and the homogeneous linewidth $\Gamma_e$ when $t_{exp}$ tends to zero.
c) i) Bare PLE spectrum. ii) PLE spectrum with three holes that have been burnt by applying a laser at the detunings $\Delta \nu=-3.1, 0.0$ and $2.2$~GHz, with respect to the inhomogeneous line center before the read-out. The solid line is a fit. 
}\label{HB3}
\end{figure}

To perform PLE, the resonant laser power and scan durations are now chosen so that there is no charge transfer during one scan.
A green laser is also turned on before and after the scan to recover the charge state SiV$^-$ and the PLE spectrum is accumulated over several such scans.
A typical PLE scan around transition C (cf. Fig.~\ref{setup}-c) is shown in Fig.~\ref{HB3}-a)
and features similar asymmetrical profiles than the ones observed in Ref.~\citenum{Nicolas2018} where a Fabry-P\'erot cavity was employed.
Here, a minimal fit to the PLE spectrum is found by using three displaced Gaussian curves of varying widths. These are the dashed lines in Fig. \ref{HB3}-a). 
Although the overall width of the PLE spectrum is found to be 7 GHz, this fit suggests that the spectrum is comprised of several emitters of varying width and central position.

Let us give a tentative interpretation to the fit that we obtain.  
Axial strain shifts the line center of the SiV$^-$ center towards to high wavelengths, whereas transverse strain shifts all the lines frequencies to the blue\cite{meesela2018}. 
Since the 4 SiV$^-$ lines feature a tail dragging to high frequencies, a possible effect is that the ensemble of SiV$^-$ transitions at the lower frequencies experience a more homogeneous axial strain. The SiV$^-$ centers that have their ZPL transitions centered 5~GHz away from the central lines may be at another more strained position in the sample, and experience a broader strain distribution.
One subgroup of SiV may be located in the original HPHT seed. The other may be in the CVD layers from the growth process. A more realistic model would include more families with widths continuously increasing towards high frequencies. 
Another explanation could be the presence of a strain gradient together with a SiV concentration gradient. The density is higher at the apex of the tip where the strain is fixed by the seed crystal. Going to the base of the pyramid, the strain would continuously change to the value specific to the CVD bulk while the concentration decreases \cite{Nelz2016}. 

Now that PLE can be efficiently performed, and that the PLE signal comes from the sum of homogeneous lines from emitters having different ZPL frequencies, one can exploit the charge state transfer to burn holes in the inhomogeneous profile and access spectral features below the inhomogeneous width. This is what we now demonstrate.

\section*{Persistent Hole Burning}

Hole burning ensembles of SiV centers was already demonstrated in ~Ref. \citenum{arend2016photoluminescence} with a resonant laser saturating one class of emitters. Here, we use the frequency dependence of the dark state trapping to demonstrate persistent hole burning.
The sequence is depicted in Fig.\ref{HB3}-b). 
After a short green illumination that ensures that the SiV centers are in the bright charged state, the sample is exposed to resonant light at a fixed frequency at the center of the inhomogeneous profile $\nu=0$ during a time $t_{\rm exp}$ and at a power of 0.877 $\mu$W. The frequency of the resonant laser is then scanned over tens of GHz around the central frequency during a time $t=100$ ms that is shorter than the transfer rate to the dark charge state at this power level. The PL count rate is recorded during the scan. To obtain a high enough signal to noise ratio, another green laser pulse is applied after the scan and the sequence is repeated. The resulting spectra are plotted in Fig~\ref{HB3}-b).
As expected, the PLE spectra now feature holes within the inhomogeneous profile, centered at a frequency $\nu=0$ and with widths that depend upon the exposure time $t_{\rm exp}$.  The lifetime of the hole is most likely determined by the lifetime of the charge states, meaning more than one hour. 
Note that this hole can be controllably undone by shining green laser light. 

To describe the dynamics of the hole burning and to have access to the homogeneous linewidth, let us now present a model based on rate equations. We here assume that the spectral distribution of emitters is given by the PLE spectrum and that they all have the same spectral linewidth $\Gamma_e$ for simplicity.

We assume that the recombination time of the electrons promoted to the conduction band is faster than all the other phenomena at play. We also assume that the rate of excitation from the excited state to the conduction band $a\Phi$ is much weaker than the excitation rate on the ZPL.
In that case, the loss rate $1/\tau_L(\nu_0)$ of SiV$^-$ centers centered at the frequency $\nu_0$ and excited by a laser at the frequency $\nu_{exc}$ with a photon flux $\Phi$ is $$\frac{1}{\tau_L(\nu_0)}=a\Phi \times \frac{b\Phi}{\frac{\Gamma_e^2}{4}+\left(\nu_0-\nu_{exc}\right)^2},$$
where the first term of the product is the excitation from the excited state to the conduction band rate, and the second term is the Lorentzian excitation rate of the SiV$^-$ on the ZPL transition. This last expression is justified by our observations that the SiV$^-$ ZPL transition is not saturated at the low red laser powers we use.
This equation predicts that the number of remaining bright emitters depends on the square of the optical power under the above assumptions and is thus compatible with a two-photon excitation process.
The overall photoluminescence rate ${\rm PLE}(\nu,t)$ under excitation at the frequency $\nu$ after exposure to the laser at frequency $\nu_{exc}$ during a time $t$, is then the sum of the contribution from all SiV$^-$ centers : 

\begin{equation}
{\rm PLE}(\nu,t)\propto\int_{-\infty}^{\infty}\frac{d(\nu_0)}{\frac{\Gamma_e^2}{4}+\left(\nu_0-\nu\right)^2}e^{-\frac{t}{\tau_L(\nu_0)}}d\nu_0,
\end{equation}
where $d(\nu_0)$ is the spectral distribution of the emitters. 

Using this model, we can now compute the PLE decay rate at fixed excitation wavelength. 
We used it to predict and to fit the change of the PL as a function of time under resonant excitation, in the experimental traces shown in Fig. 3-a). 

We also used this model to fit the hole burning spectra presented in Fig. 4 and use the fitted PLE spectrum without the hole burning as the spectral density $d(\nu_0)$.
To accurately estimate the homogeneous broadening of our sample, we acquire several hole burning spectra at the same laser power, but at different exposure times. 
We then use three parameters: $C_1=\Gamma_e$, $C_2=\Phi^2 a b t$ and $C_3=\nu_{exc}$ and fit all the four spectra of the series at the same time. We thus have a set of six parameters to optimize: $C_1$, $C_3$ and four $C_2$ values. To check the validity of this procedure, we verify that the parameter $C_2$ is a linearly function of the exposure time.
This leads to an estimation of $\Gamma_e$ of $390\pm 12$ MHz.

Importantly, according to this model, the linewidth of the hole cannot be lower than twice the homogeneous linewidth. 
This can be seen by assuming a small interrogation time and doing the Taylor expansion
\begin{equation}
\label{eq1}
{\rm PLE}(\nu,t)=A+B \int_{-\infty}^{\infty}\frac{1}{\frac{\Gamma_e^2}{4}+\left(\nu_0-\nu\right)^2}\frac{1}{\frac{\Gamma_e^2}{4}+\left(\nu_0-\nu_{exc}\right)^2}d\nu_0
\end{equation}
where A and B are independent on frequency. This corresponds to the convolution of two Lorentzian functions. The result is then a Lorentzian function with a width that is twice as large as the initial Lorentzian widths.

This charge state conversion therefore allows us to perform efficient persistent spectral hole burning and to have access to the homogeneous linewidth. Previous studies on the same pyramids have measured the lifetime of the excited state to be 0.81 ns which corresponds to a lifetime-limited linewidth of 196 MHz \cite{Nelz2016}. The measured homogeneous linewidth is thus only two times greater here, which is very rare for ensembles in a nanostructure.

Further work will be conducted to estimate the contribution of spectral diffusion and temperature on the obtained width, but our results already show that it is very small in such nano-pyramids, which opens perspectives for preparing narrow persistent spectral features containing many SiV$^-$ centers. 

The long lifetime of the holes also enables us to prepare more complex spectral patterns. 
This is demonstrated in Fig. \ref{HB3}, where three holes are burnt consecutively by applying the resonant laser at three different frequencies within the inhomogeneous profile. 
The resulting spectrum shows the versatility of the charge-state mediated persistent hole burning and opens a path towards the preparation of narrow anti-holes and to more exotic spectral features that could be employed for light storage protocols, for instance using the atomic frequency comb technique \cite{Afzelius2009}. 

\section*{Conclusion}

In conclusion, we have demonstrated close to lifetime-limited emission from a high density ensemble of SiV centers in a diamond nano-pyramid at cryogenic temperature. 
We have done so using resonant photoluminescence excitation in conjunction with charged state transfer, which induces persistent hole burning on samples where nitrogen impurities are present. 
We have also shown that these impurities are required to trap either electrons or holes in order to enable the charge state manipulation. 

At room temperature, this effect could also lead to applications for biological imaging. Combining ground state depletion microscopy (GSD) using the charge state conversion (already performed with NV centers \cite{NVGSD}) 
and diamond bio-imaging techniques \cite{Claveau2018} could indeed enable sub-wavelength resolution. 
The hole burning that we show is unique in that it is fatigue free and is all optically controlled via a long lived dark states.
At present the lifetime of the ground orbital states is most likely limited both by the coupling to impurities and by photons. Using spin echoes and using a dilution fridge to reach mK temperatures could then enable SiV ensembles to be used as an efficient quantum memory for light with long coherence times.\\



We would like to thank Elke Neu, Vincent Jacques and the nano-optics team at LPENS for fruitful discussions and Romaric Le Goff for technical assistance.  GH acknowledges funding by the French National Research Agency (ANR) through the project SMEQUI and by the T-ERC program through the project QUOVADIS. 


\providecommand{\latin}[1]{#1}
\providecommand*\mcitethebibliography{\thebibliography}
\csname @ifundefined\endcsname{endmcitethebibliography}
  {\let\endmcitethebibliography\endthebibliography}{}

\end{document}